\documentclass[aps,pre,onecolumn,showpacs,superscriptaddress]{revtex4-2}
\usepackage{graphicx}
\usepackage{dcolumn}
\usepackage{bm}
\usepackage{amsmath,amssymb}
\usepackage{mathrsfs}
\usepackage{color}
\usepackage{makeidx}

\usepackage{subcaption} 
\makeindex

\usepackage{float}
\makeatletter
\let\newfloat\newfloat@ltx
\makeatother

\usepackage{algorithm}
\usepackage{algpseudocode}
\usepackage{comment}

\newcommand{\dist}{\operatorname{dist}}

\begin{document}

\title{Structural robustness of networks with degree-degree correlations between second-nearest neighbors}

\author{Yuka Fujiki}\email{yfujiki@tohoku.ac.jp}
\affiliation{Frontier Research Institute for Interdisciplinary Sciences (FRIS), Tohoku University, 6-3 Aramaki aza Aoba, Aoba-ku, Sendai 980-8578, Japan}
\affiliation{Advanced Institute for Materials Research (AIMR), Tohoku University, 2-1-1 Katahira, Aoba-ku, Sendai, 980-8577 Japan}
\author{Stefan Junk}\email{sjunk@math.gakushuin.ac.jp}
\affiliation{Gakushuin University, 1-5-1 Mejiro, Toshima-ku, Tokyo 171-8588 Japan}

\date{\today}
\begin{abstract}
We numerically investigate the robustness of networks with degree-degree correlations between nodes separated by distance $l=2$ in terms of shortest path length.
The degree-degree correlation between the $l$-th nearest neighbors can be quantified by Pearson's correlation coefficient $r_l$ for the degrees of two nodes at distance $l$.
We introduce $l$-th nearest-neighbor correlated random networks ($l$-NNCRNs) that are degree-degree correlated at less than or equal to the $l$-th nearest neighbor scale and maximally random at farther scales.
We generate $2$-NNCRNs with various $r_1$ and $r_2$ using two steps of random edge rewiring based on the Metropolis-Hastings algorithm and compare their robustness against failures of nodes and edges.
As typical cases of homogeneous and heterogeneous degree distributions, we adopted Poisson and power law distributions.
Our results show that the range of $r_2$ differs depending on the degree distribution and the value of $r_1$. Moreover, comparing $2$-NNCRNs sharing the same degree distribution and $r_1$, we demonstrate that a higher $r_2$ makes a network more robust against random node/edge failures as well as degree-based targeted attacks, regardless of whether $r_1$ is positive or negative.
\end{abstract}
\maketitle

\section{Introduction}


Many complex networks in the real world share a common property that the distribution of the number of edges (degree) from each node is approximated by a power-law function, which is known as the scale-free property \cite{Barabasi1999}.
The degree correlation among nodes in such networks, particularly nearest-neighbor degree correlation (NNDC), influences network behavior, including robustness to node and edge failures, the spread of infections, and oscillator synchronization \cite{Newman2002assr, Dorogovtsev2008, Pastor-Satorras2015}.

\smallskip Additionally, recent studies have shown that the majority of complex networks in the real world exhibit long-range degree correlations (LRDCs) at greater distances than next-nearest neighbors \cite{Fujiki2018}.
While some LRDCs can be explained as ``extrinsic" LRDCs resulting from the propagation of the NNDCs, for the most part, these are ``intrinsic" LRDCs that are not merely consequences of the NNDCs \cite{Fujiki2020}.
The effects of LRDCs remain poorly understood, except for some special cases.
For example, it is known that the shortest path length between high-degree nodes influences network functions and dynamical properties \cite{Tadic2004,Rybski2010,Boguna2013,Boulos2013,Swanson2016,Fujiki2017}, and that transsortative structures (degree correlations between two neighbors of a node) can amplify the majority illusion \cite{Ngo2020}.
Towards understanding the relationship between network properties and LRDCs, in this study, we numerically investigate the influence of LRDC at shortest path distance $l=2$ on the structural robustness, which is one of the most fundamental properties of networks.

\smallskip Robustness refers to the ability of a network to maintain its overall structural and functional integrity against node or edge removal \cite{Newman2003rev,Albert2002,Cohen2010,Liu2022,Artime2024}.
This concept may be interpreted as assessing the stability of network functioning, or the difficulty of eradicating infectious diseases and misinformation spreading on the network.
Taking only the degree distribution of a network into account, analytical calculations using the mean-field approximation show that networks with homogeneous degree distributions lose global connectivity earlier, i.e., after a smaller percentage of random failures, than heterogeneous degree distributions \cite{Molloy1995,Callaway2000}.
On the other hand, it is known that heterogeneous networks are more vulnerable to targeted attacks than homogeneous ones \cite{Albert2001,Cohen2001}.
Beyond the effect of the degree distribution, it is known that positive, resp. negative, NNDCs makes the network more robust, resp. weaker,~\cite{Goltsev2008,Shiraki2010,Herrmann2011,Tanizawa2012}.
If LRDCs influence robustness in a manner similar to NNDCs, this understanding would provide valuable insights into the resilience of networks and contribute to more effective system design.

\smallskip For this purpose, we introduce $l$-th nearest-neighbor correlated random networks ($l$-NNCRNs) that are degree-degree correlated up to the $l$-th nearest-neighbor distance and maximally random at any further scale.
In particular, the difference between $l$-NNCRNs and $(l-1)$-NNCRNs is caused by intrinsic LRDCs at distance $l$.

\smallskip Formally, LRDCs at distance $l$ can be characterized by the conditional probability $P(k,k'|l)$ that the degrees of two nodes are $k$ and $k'$ given that the two nodes are at distance $l$ from each other in terms of the shortest path length, see Sec.~\ref{sec:defs} below.
Thus, to isolate the effect of LRDCs at a distance $l$, we can compare properties of $l$-NNCRNs to networks with the same correlations $P(k,k' |l')$ at distances $l'\leq l-1$ and no intrinsic correlations at distance $l$. The strength of degree correlations at distance $l$ is quantified by Pearson's correlation coefficient $r_l$ for $P(k,k' |l)$.

\smallskip We numerically implement this framework in the case $l=2$ for a representative choice of correlations.
Our results demonstrate, firstly, that the correlations at distances $l=0$ and $l=1$, i.e. the degree distribution and $r_1$, place some constraints on the range of $r_2$.
We then perform random node or edge removal and degree-based targeted attack on networks and quantify the robustness based on the percolation critical point $p_c$ and the robustness measure $R$ introduced by Schneider {\it et al.}~\cite{Schneider2011}.

\smallskip Qualitatively, our results suggest that the critical point $p_c$ shifts with $r_2$ when the correlations at shorter distances are kept fixed, which proves that LRDCs indeed influence network robustness. The effect of $r_2$ on robustness follows a similar trend to that of $r_1$. 
Networks with larger $r_2$ tend to be more robust against both random node/edge failure and degree-based targeted attacks when the robustness is quantified by $f_c=1-p_c$. Using another measure of robustness, $R$, networks with larger $r_2$ are more fragile against random node/edge failure but remain more robust against degree-based targeted attacks. The difference is because both measures quantify slightly different aspects of robustness, as will be explained in Sec~\ref{randfail}.
Quantitatively, the influence of $r_2$ on network robustness is weaker than that of $r_1$, but it is generally on a comparable order.
To clarify how degree correlations at distance $l=2$ affect network robustness, we decompose $2$-NNCRNs into $k$-cores and confirm that a strongly positive $r_2$ induces a core, which is as robust as the core induced by $r_1$.
We close the discussion by investigating the values of $r_2$ exhibited by real networks and the possibility for optimization of robustness based on LRDCs.

\smallskip The manuscript is organized as follows.
In Sec.~\ref{sec:method} we define $l$-NNCRNs by quantifying LRDCs using $P(k,k'|l)$ and $r_l$.
Moreover, we explain the algorithm for generating $2$-NNCRNs with various values of $r_l$ using random edge rewiring based on the Metropolis-Hasting algorithm.
Sec.~\ref{sec:result} presents the numerical results obtained from our simulations, including the effects of LRDCs on network robustness and the relationship between $r_1$, $r_2$ and the degree distribution.
Finally, in Sec.~\ref{sec:end} we summarize our key findings and propose avenues for future research in elucidating the role of LRDCs in complex networks.

\section{Numerical method}\label{sec:method}
We numerically sample networks with various degree-degree correlations between nodes at distance $l\leq 2$ in terms of shortest path length.


\smallskip We propose $l$-th nearest neighbor correlated networks ($l$-NNCRNs), which are random at shortest path distances greater than $l$ and which serve as an idealized model to study the relation between network robustness and LRDCs without other confounding factors.
$2$-NNCRNs are sampled using a three-step edge-rewiring algorithm to fix, successively, the degree sequence, the NNDCs and finally the LRDCs at distance $l=2$. 
The definition and sampling method of $l$-NNCRNs are explained in the following two subsections.

\smallskip Note that in real-world networks, the effect of $r_2$ is complicated by the fact that $r_1$ by itself also influences network robustness and, on the other hand, constrains the range of $r_2$.
In addition, intrinsic long-range correlations at distances $l \geq 3$ may exist and affect robustness. $2$-NNCRNs do not have such intrinsic long-range correlations at distances $l \geq 3$, allowing us to isolate the effects of $r_1$ and $r_2$.
Moreover, by preparing $2$-NNCRNs with identical $r_1$ but varying $r_2$, we can focus specifically on the effect of $r_2$. This idealized model enables us to disentangle these effect and to demonstrate that LRDCs at distance $l=2$ can have a profound effect on network robustness.
To investigate the relation between the robustness and LRDCs at $l=2$, we compare the effect of random node/edge removal and of degree-based targeted attack on the network structure in $2$-NNCRNs with different degree correlations.
The analysis of these properties is explained in the final subsection.

\subsection{Definition of $l$-NNCRNs}\label{sec:defs}
To analyze LRDCs in a given network, we first need to understand maximally random networks that have the same nearest-neighbor correlations as the network \cite{Fujiki2020}.
We can use that model as a baseline to judge whether the degree correlations at greater distances than $l=1$ are just extrinsically caused by the NNDCs or whether they are intrinsic.
Extending this, we call a network ensemble $l$-NNCRN if it is degree-degree correlated at less than or equal to the $l$-th nearest neighbor distance and maximally random at any further scale.
A nearest-neighbor correlated random network is thus a $1$-NNCRN.

\smallskip The conditional probability $P(k,k'|l)$ that two randomly chosen nodes separated at a distance $l$ have degree $k$ and $k'$ describes the LRDCs at distance $l$.
Since the conditional probability is a high-dimensional matrix and not easy to interpret, we use the Pearson's correlation coefficient $r_l$ for the degrees of two nodes at distance $l$ as a convenient observable that quantifies the strength of the $l$-th nearest neighbor degree correlation.
It is defined as
%
\begin{align}
    r_l&=\displaystyle\frac
    {\sum_{k,k'}{kk'P(k,k'|l)}-[\sum_{k,k'}{kP(k,k'|l)}]^2}
    {\sum_{k,k'}{k^2 P(k,k'|l)}-[\sum_{k,k'}{kP(k,k'|l)}]^2}\in[-1,1],
\end{align}
where the sum is over all possible degrees.
It can also be represented as
\begin{align}
    r_l&=\displaystyle\frac
    {\frac{1}{{2M}^{(l)}} \sum_{i,j}{\delta_{\dist(i,j),l}\,  k_i k_j }
    - \left[ \frac{1}{{2M}^{(l)}} \sum_{i,j} { \delta_{\dist(i,j),l}\, k_i } \right]^2}
    {\frac{1}{{2M}^{(l)}}\sum_{i,j}{\delta_{\dist(i,j),l}\,  k_i^2}
    - \left[ \frac{1}{{2M}^{(l)}} \sum_{i,j} { \delta_{\dist(i,j),l}\, k_i} \right]^2}.
\end{align}
Here, the sum is over all nodes, $\dist(i,j)$ is the shortest path distance between $i$ and $j$.
The term $M^{(l)}$ is defined as
\begin{align}
M^{(l)}=\frac 12\sum_{i,j} {\delta_{\dist(i,j),l}},
\end{align}
where $\delta_{d,l}$ is Kronecker $\delta$.
In the case of $l=1$, this quantity is known as assortativity \cite{Newman2002assr} and it has been extended to arbitrary $l>1$ by \cite{Mayo2015,Arcagni2017}.

\smallskip Note that the NNDCs can induce extrinsic LRDCs at distance $l=2$, so that the value of $r_2$ may differ from zero even if the network is a $1$-NNCRN.
For a given network, we thus define $r_2^{\operatorname{ext}}$ to be the average value of $r_2$ among networks with the same degree sequence and NNDC, as described by $P(k,k'|l=1)$.
We say that a network is intrinsically degree-degree correlated at $l=2$ if the value of $r_2$ significantly differs from $r_2^{\operatorname{ext}}$.

\subsection{Edge rewiring algorithm}\label{sec:algo}
Edge rewiring algorithms are widely used to investigate graph structures \cite{Xulvi-Brunet2004,Noh2007,Orsini2015}.
Through randomization via rewiring, the algorithm can generate the most randomized network ensembles under specific constraints.
In this study, given parameters $J_1$ and $J_2$, we generate $2$-NNCRNs through the following sequence of steps based on the Metropolis-Hasting algorithm started with an initial network $G_0$:
(1) Rewire edges while preserving the degree sequence to generate a $1$-NNCRN $G_1$(see Algorithm~1).
(2) Rewire edges in the resulting network while preserving the NNDC (i.e., $P(k,k'|l=1)$) to generate a $2$-NNCRN $G_2$ (see Algorithm~2). 

\smallskip Note that when choosing edges, we (temporarily) equip each edge with an orientation. We also mention that the rewiring in Algorithm~1 is irreducible for the set of graphs with the same degree sequence and that in Algorithm~2 is as well for the NNDC \cite{Taylor1982,Stanton2012}. Moreover, the reason why we include Step~2 in both algorithms it to ensure that the Markov chains are aperiodic and thus converge to equilibrium.

\begin{algorithm}[b]
\caption{Generating a $1$-NNCRN while preserving the degree sequence}
\begin{algorithmic}[1] 
\State Increase time step by $1$.
\State Return to step $1$ with probability $1/2$.
 \State Randomly choose two edges $(u_1, u_2)$ and $(v_1, v_2)$.
 \State If rewiring $(u_1, u_2)\to(u_1,v_2)$ and $(v_1, v_2)\to(v_1,u_2)$ creates a loop or a multi-edge, return to step $1$.
 \State Compute the current value $r_1^{prev}$ of $r_1$, the value $r_1^{next}$ of $r_1$ after rewiring $(u_1, u_2)\to(u_1,v_2)$ and $(v_1, v_2)\to(v_1,u_2)$, and the transition probability $P(u_1,u_2,v_1,v_2) = \min(1,\exp[-J_1 (r_1^{next}-r_1^{prev} )])$.
 \State With probability $P(u_1,u_2,v_1,v_2)$, rewire edges $(u_1, u_2)\to(u_1,v_2)$ and $(v_1, v_2)\to(v_1,u_2)$.
 \State Repeat steps 1--6 until the system reaches equilibrium.
\end{algorithmic}
\end{algorithm}

\begin{algorithm}[t]
\caption{Generating a $2$-NNCRN while preserving the degree sequence and $P(k,k'|l=1)$}
\begin{algorithmic}[1] 
\State Increase time step by $1$.
\State Return to step $1$ with probability $1/2$.
\State Randomly choose an edge ($u_1$, $u_2$).
\State Randomly choose another edge ($v_1$, $v_2$) such that $v_1$ has the same degree as $u_1$.
\State If rewiring $(u_1, u_2)\to(u_1,v_2)$ and $(v_1, v_2)\to(v_1,u_2)$ creates a loop or a multi-edge, return to step $1$.
\State Compute the current value $r_2^{prev}$ of $r_2$, the value $r_2^{next}$ of $r_2$ after rewiring $(u_1, u_2)\to(u_1,v_2)$ and $(v_1, v_2)\to(v_1,u_2)$, and the transition probability $P(u_1,u_2,v_1,v_2) = \min(1,\exp[-J_2 (r_2^{next}-r_2^{prev} )])$.
\State With probability $P(u_1,u_2,v_1,v_2)$, rewire edges $(u_1, u_2)\to(u_1,v_2)$ and $(v_1, v_2)\to(v_1,u_2)$.
\State Repeat 1--7 until the system reaches equilibrium.
\end{algorithmic}
\end{algorithm}

\smallskip These algorithms generate a canonical ensemble of networks in the sense of statistical physics, which satisfies a constraint in terms of the mean value (soft constraint) and belongs to the general class of exponential random graph models \cite{Berg2002,Park2004,Cimini2019}.

\smallskip More precisely, in Algorithm~$1$ the degree of each node does not change, so the degree sequence acts as a hard constraint.
On the other hand, the rewiring generates a set of networks with different NNDCs but which is random at further distances, i.e., a $1$-NNCRNs.
The mean value of $r_1$ is a soft constraint in Algorithm~1 and in equilibrium a configuration $G_{1}$ (with the same degree sequence as the initial network $G_{0}$) appears with probability
\begin{align}
    \Pi_1(G_{1}|G_{0})=\frac{e^{-J_1 r_1(G_{1})}}{Z_1(J_1|G_{0})},
\end{align}
where $r_1(G)$ is the value of $r_1$ for $G$ and $Z_1$ is the partition function
\begin{align}
    Z_1(J_1|G_{0})=\sum_{G'}{e^{-J_1 r_1(G')}}.
\end{align}
The mean value of $r_1$ is zero in the case $J_1=0$ and moves in the positive/negative direction with $J_1$. Moreover, when the network size is sufficiently large, the value of $r_1$ tends to be narrowly distributed around its mean value (notice that the error bars in Fig~\ref{fig:1} are barely visible).

\smallskip The idea behind Algorithm~$2$ is similar, but in addition to the degree sequence the condition that $u_1$ and $v_1$ have the same degree ensures that $P(k,k'|l=1)$ is preserved, and in particular that $r_1$ remains constant. 
When the system reaches equilibrium in Algorithm~2, a configuration $G_{2}$ with the same degree sequence and NNDC as $G_1$ appears with probability
\begin{align}
    \Pi_2 (G_{2}|G_{1})=\frac{e^{-J_2 r_2(G_{2})}}{Z_2 (J_2|G_{1} )},
\end{align}
where $r_2(G)$ is the value of $r_2$ for $G$ and $Z_2(J_2|G_1)$ is the partition function in the form of
\begin{align}
    Z_2(J_2|G_{1})=
    \sum_{G'}{e^{-J_2 r_2(G')} }.
\end{align}
The mean value of $r_2$ equals $r_2^{\operatorname{ext}}$ in the case $J_2=0$ and moves in the positive/negative direction in accordance with $J_2$.
Both rewiring procedures satisfy the detailed balance condition. Thus, after successively applying Algorithms $1$ and $2$ to an initial network $G_{0}$ until equilibrium is reached, we obtain network $G_{2}$ with probability
\begin{align}
    \sum_{G_{1}}\Pi_1(G_{1}|G_{0})\Pi_2(G_{2}|G_{1}).
\end{align}

We prepare two kinds of random networks as the initial network $G_0$, which prescribes the degree distribution of the network and which is preserved by Algorithms~1 and 2.
The first is the Erd\H{o}s-R\'enyi random graph model, whose degree distribution is approximated by a Poisson distribution $P(k)=\langle{k}\rangle^k e^{-k}/k!$ when the network is large enough.
We set the number of nodes and the average degree at $N=20,000$ and $\langle{k}\rangle=5.0$, respectively.
The second is the configuration model \cite{Newman2001cm} with a power-law degree distribution $P(k)=ck^{-\gamma}$ with the number of nodes $N=20,000$, the power-law exponent $\gamma=2.5$, the minimum degree $k_{min}=2$, and the structural cutoff $k_{c}=\sqrt{N}$~\cite{Boguna2004}, where $c$ is the normalization constant.

\smallskip While the former network belongs to the group of networks with homogeneous degree distribution, the later is a network with power-law degree distribution, which is one of the common features shared by many real-world complex networks.
Empirically, power law exponents in the range $2\lesssim \gamma\lesssim 4$ are common.
Such a heterogeneous degree distribution tends to amplify the effect of degree-degree correlations on robustness and other properties of networks.
Hereafter, we refer to these networks as Poisson and power-law networks, respectively.

\subsection{Analyzing the structural robustness}\label{sec:rob}
The structural robustness of networks assesses the ease with which the global connectivity of the network is maintained.
There are several strategies for destroying networks and processes of how networks fail \cite{Callaway2000,Albert2001,Cohen2001,Holme2002}.
Here, we investigate the most straightforward mechanisms, namely random node/edge failures, as well as targeted attacks removing nodes in decreasing order of degree.

\smallskip As the fraction $p$ of remaining nodes/edges decreases, the size of the largest connected component $N_{\operatorname{LCC}}(p)$ in the network decreases as well.
For the network to maintain its functionality, global connectivity is necessary.
Thus the value of $p$ at which the largest connected component collapses, the critical point $p_c$, serves as an essential indicator of robustness.
We find it more intuitive to evaluate robustness using the fraction of removed nodes/edges required for collapse, $f_c = 1 - p_c$, so that a larger value of $f_c$ indicates that the network is more robust.

\smallskip In this study, the network size is fixed at $N = 20,000$ and we approximate the critical point $p_c$ using a threshold of $p_{0.005}=\sup\{p\colon\langle N_{\operatorname{LCC}}(p)\rangle< 0.005N\}$ where the giant component size reaches $0.5$\% of the total to approximate the critical point $p_c$.
Moreover, we use $f_{0.005}=1-p_{0.005}$ as a robustness measure.

\smallskip In practice, $N_{\operatorname{LCC}}(p)$ does not always rapidly decrease around a single value of $p$. More precisely, the network may contain a densely connected core whose percolation occurs later than the more loosely connected periphery, which results in multiple phase transitions. The threshold $p_{0.005}$ has been chosen to detect the ``true'' collapse of network connectivity where all connected components become sublinear in the network size $N$. To illustrate this phenomenon, we compute the susceptibility
\begin{align}
    \chi(p)=\frac{\langle{N_{\operatorname{LCC}}^2(p)}\rangle-\langle{N_{\operatorname{LCC}}(p)}\rangle^2}{\langle{N_{\operatorname{LCC}}(p)}\rangle}.
    \label{eq:chi}
\end{align}
The peak of the susceptibility suggests the location of critical points in a continuous phase transition and is often used to approximate $p_c$.
However, in situations as described above the susceptibility exhibits multiple peaks how relative sizes can flip as the network size increases, which makes it unsuitable for quantifying robustness.
The multiple peaks phenomenon can be seen in the bottom right panel in Fig.~\ref{fig:3}.

\smallskip Beyond the critical point, the size of the giant component as $p$ varies also carries important information. Schneider {\it et al.} \cite{Schneider2011} proposed the area $R=\int_0^1 S(p)\text{d} p$ under the curve $S(p)=N_{\operatorname{LCC}}(p)/N$ as a robustness measure. A larger value of $S(p)$ indicates that the network is more robust for any fixed fraction $p$ of remaining nodes/edges and $R$ thus measures the average robustness away from $p_c$.

\smallskip In summary, network robustness can be interpreted using two measures: $f_{0.005}$ and $R$. A larger value of $f_{0.005}$ indicates that the network can maintain its global connectivity under a higher fraction of node/edge removals. The robustness measure $R$, defined as the area under the curve $S(p)$, reflects the average connectivity across all node/edge removal scenarios, with a larger $R$ indicating greater robustness.

\smallskip To evaluate $f_{0.005}$ and $R$ in a given network, we use the method proposed by Newman and Ziff \cite{Newman2001ziff}.

\section{Result}\label{sec:result}
In this section, we present the results of the numerical calculations based on the method described in Sec.~\ref{sec:algo}.
First, we confirm that the algorithm works as desired, i.e., that the generated $2$-NNCRNs exhibit various degree-degree correlations between both nearest-neighbor nodes and second-nearest-neighbor nodes.
Next, we assess their robustness against random node/edge failures and degree-based targeted attacks.

\subsection{Constraints on degree correlations in $2$-NNCRNs}\label{r2range}
The purpose of this section is to give some details about the networks obtained by the rewiring procedure explained above, and in particular to explain how the parameters influence the range of correlations achievable with our method.
More precisely, we confirm that degree distribution and $r_1$ impose constraints to $r_2$ and  that the rewiring algorithm can sample network configurations with various $r_2$ and the same $r_1$.
 \begin{figure}[tbtbtb]
    \begin{center}
     \includegraphics[width=0.7\linewidth]{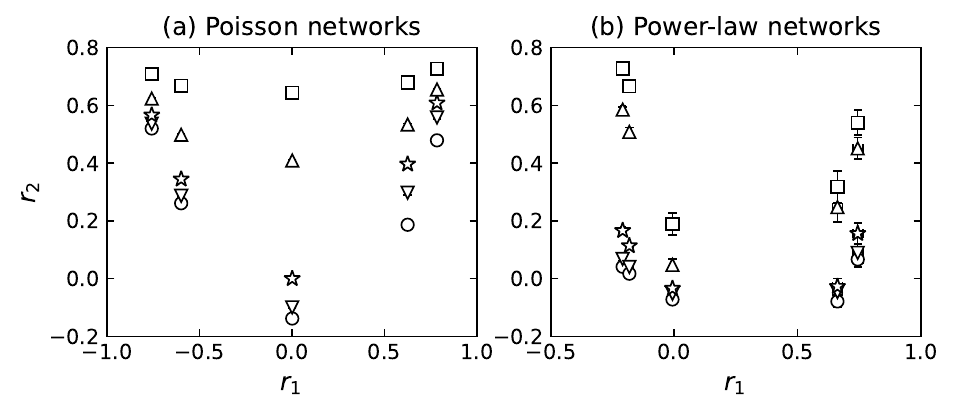}
    \caption{
    The mean value of $r_1$ and $r_2$ of $2$-NNCRNs.
    Each data point is the average value over $100$ configurations, and the error bar is the standard deviation.
    The $r_1$ and $r_2$ of power-law networks exhibit larger error bars than those of Poisson networks, which can be attributed to significant differences in maximum degrees across random seeds.
    The symbols vary based on the rewiring parameter $J_2$.
    For $J_2 = -2M, -M, 0, M, 2M$, the symbols are a sphere, a downward triangle, a star, an upward triangle, and a box, respectively, where $M$ is the number of edges in each network.
    In particular, the star indicates $J_2 = 0$, where the value of $r_2^{\operatorname{ext}}$ reflects the extrinsic LRDC at $l=2$ induced by $r_1$ in $1$-NNCRNs.
    If $r_2$ is greater than/less than $r_2^{\operatorname{ext}}$, it implies that an intrinsic positive/negative LRDC is present at $l=2$.
    }
     \label{fig:1}
     \end{center}
 \end{figure}

\smallskip
Figure~\ref{fig:1} depicts the average values of $r_1$ and $r_2$ of $2$-NNCRNs generated by the rewiring method described in the previous section starting with an initial network characterized by (a) the Poisson degree distribution and (b) the Power-law degree distribution. As shown in Fig~\ref{fig:1}, the range of $r_2$ strongly depends on the degree distribution and $r_1$. Let us comment on a few trends:

\smallskip First, we note that $r_2$ is shifted in the positive direction by larger absolute values of $r_1$.
Such positive extrinsic LRDCs have also been observed in NNCRNs in \cite{Fujiki2020} and \cite{Mizutaka2016} and are consistent with the positive correlation of degrees between nodes at even distances.
To understand this effect, note that if $r_1$ is close to $+1$, similar degrees tend to be adjacent, leading to similar degrees at $l=2$ and resulting in higher values of $r_2$.
On the other hand, if $r_1$ is close to $-1$, high degrees tend to be adjacent to low degrees and vice versa. Due to this alternating pattern, we again find that nodes at distance $l=2$ have similar degrees, which results in high values of $r_2$.

\smallskip Next, we observe that the range of $r_2$ is narrower if the absolute value of $r_1$ is large in the case of Poisson degree distributed networks (Fig~\ref{fig:1}(a)).
For power-law networks, the converse is true but the effect is much weaker (Fig~\ref{fig:1}(b)).
In other words, the range of $r_2$ values achievable by $2$-NNCRNs strongly depends on the value of $r_1$ in Poisson networks, while the same trend is not observed for power-law networks.
At present, we do not have a satisfying explanation for this behavior and we conclude that the relationship between $r_1$ and $r_2$ is complex and constrained by the degree distribution.

\smallskip Finally, we mention that neither the range of $r_1$ nor of $r_2$ are symmetric.
This is not surprising -- for example, as explained above, both $r_1$ close to $+1$ and $r_1$ close to $-1$ tend to induce large positive values for $r_2$, but the mechanism is different in both cases.

\subsection{Random failure}\label{randfail}
Having confirmed our samples to be degree-degree correlated at distances $l=1$ and $l=2$ in various ways, we now investigate their robustness using the method described in Sec.~\ref{sec:rob}. 

\begin{figure}[tbtbtbtbtb]
    \centering
     \includegraphics[trim={0cm 0cm 0cm .7cm}, clip, width=1.0\linewidth]{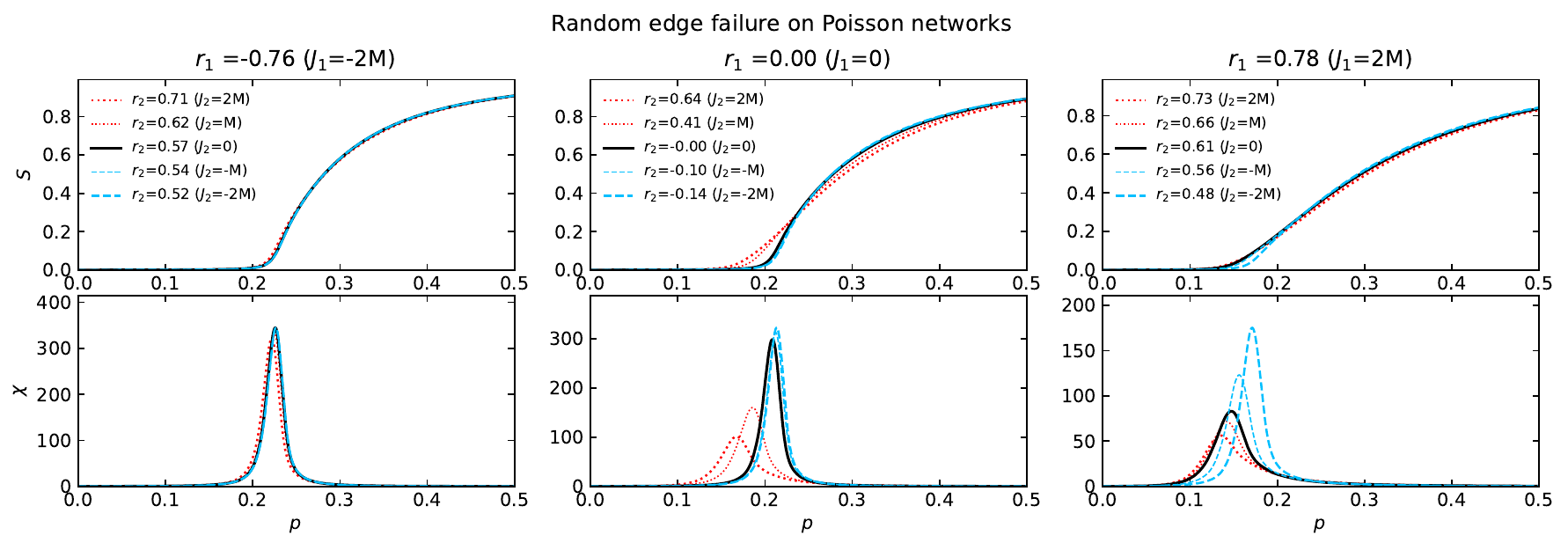}
    \caption{The $S$ and $\chi$ dependence on the edge remaining probability $p$ during random edge failure on $2$-NNCRNs with Poisson degree distribution.
    The set of $2$-NNCRNs is the same as in Fig~\ref{fig:1}(a).
    The left, middle, and right panels correspond to rewiring parameters $J_1=-2M$, $0$, and $2M$, respectively.
    The colors and types of the lines correspond to rewiring parameters $J_2$.}
    \label{fig:2}
\end{figure}

\begin{figure}[tbtbtbtbtb]
     \centering
     \includegraphics[trim={0cm 0cm 0cm .7cm}, clip, width=1.0\linewidth]{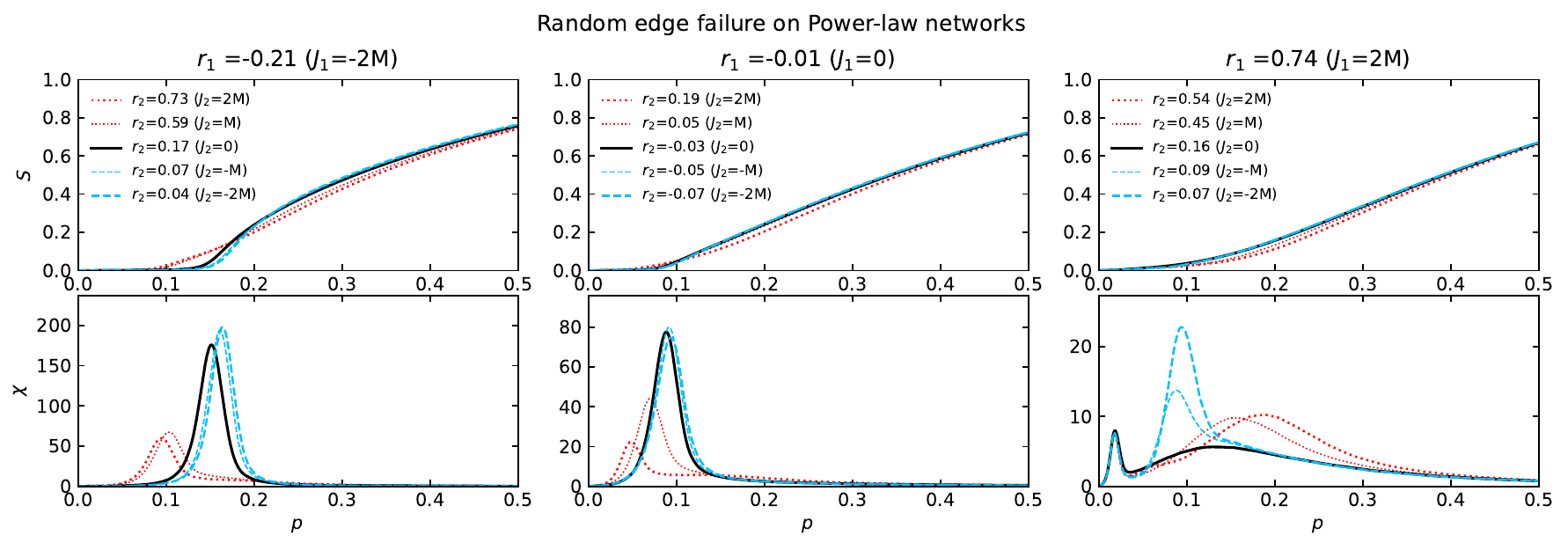}
    \caption{Here we plot the same information as Fig~\ref{fig:2} for the set of $2$-NNCRNs from Fig~\ref{fig:1}(b).}
    \label{fig:3}
\end{figure}

\smallskip Figs~\ref{fig:2}--\ref{fig:3} depict the relative sizes of the largest connected component $S=\langle{N_{\operatorname{LCC}} }\rangle/N$ and the susceptibility $\chi$ (see Eq.~(\ref{eq:chi})) as functions of the probability $p$ that an edge remains.
\begin{figure}[h]
    \centering
    \begin{subfigure}[b]{0.4\linewidth}
        \centering
        \includegraphics[trim={0cm 0cm 0cm 1.3cm}, clip, width=\linewidth]{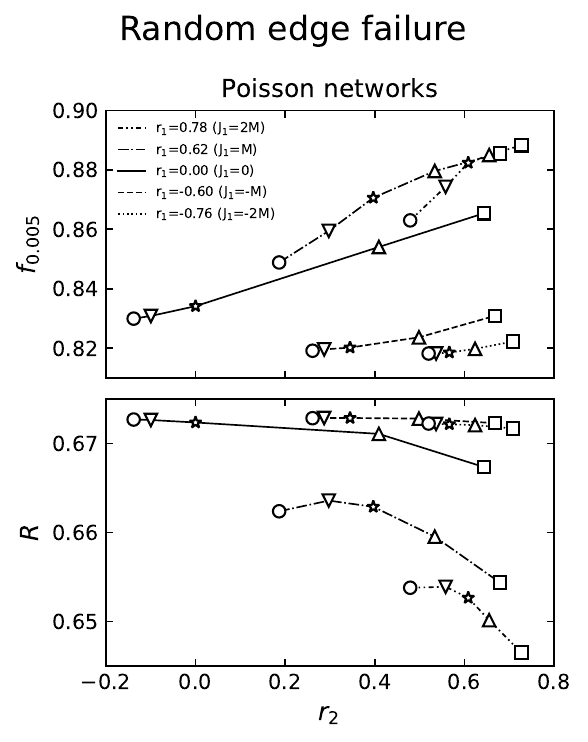}
    \end{subfigure}%
    \begin{subfigure}[b]{0.4\linewidth}
        \centering
        \includegraphics[trim={0cm 0cm 0cm 1.3cm}, clip, width=\linewidth]{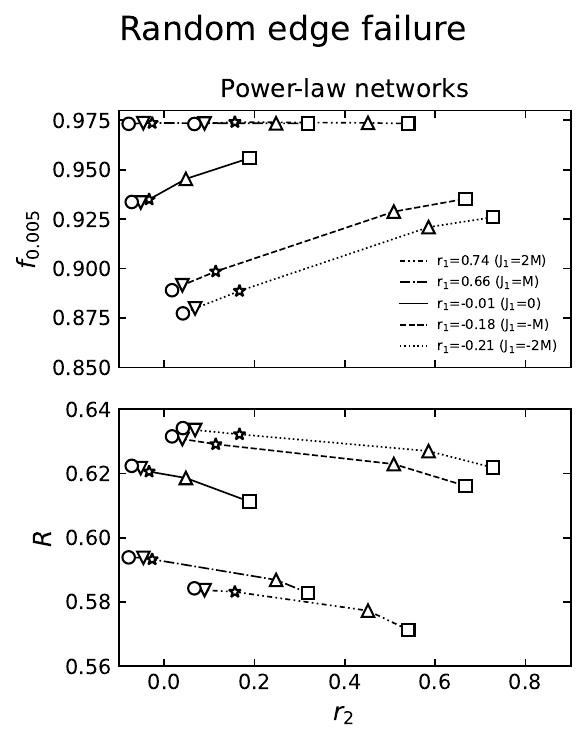}
    \end{subfigure}
    \caption{
    Robustness of $2$-NNCRNs against random edge removal.
    The symbols indicate the parameter $J_2$ and are the same as those in Fig~\ref{fig:1}, whereas lines connect points corresponding to $2$-NNCRNs with the same $r_1$ and the line type corresponds to $J_1$.
    We note that the change in in response to varying $r_2$ (the vertical displacement between the endpoints of a line) is comparable to the response to varying $r_1$ (the vertical distance between a pair of lines).
    }
    \label{fig:4}
\end{figure}
\begin{figure}[h]
    \centering
    \begin{subfigure}[b]{0.4\linewidth}
        \centering
        \includegraphics[trim={0cm 0cm 0cm 1.3cm}, clip, width=\linewidth]{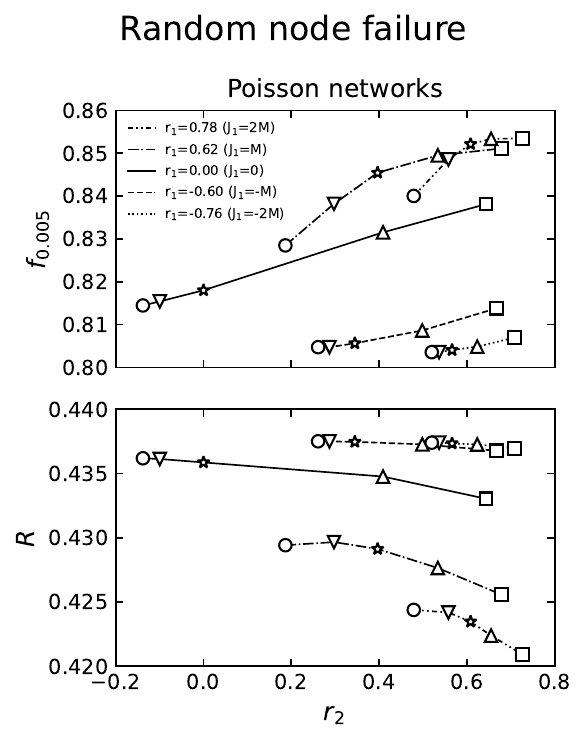}
    \end{subfigure}%
    \begin{subfigure}[b]{0.4\linewidth}
        \centering
        \includegraphics[trim={0cm 0cm 0cm 1.3cm}, clip, width=\linewidth]{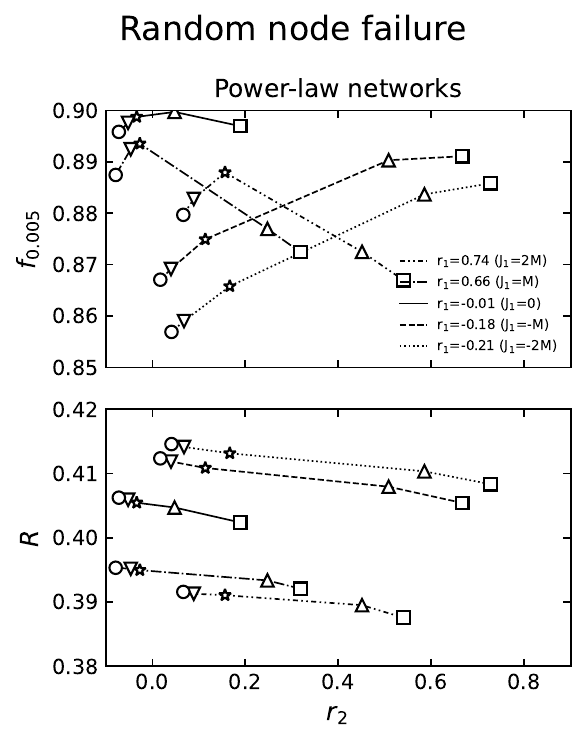}
    \end{subfigure}
    \caption{
    Robustness of $2$-NNCRNs against random node removal, analogously to Fig~\ref{fig:4}.
    It is notable that the behavior of $f_{0.005}$ in the top-right panel does not match what has been observed in the case of edge failure or targeted attacks.
    We believe this is merely an artifact of our choice to approximate $p_c$ by $p_{0.005}$, which is somewhat unstable in cases where the critical point is close to zero.  }
    \label{fig:4_site}
\end{figure}

\smallskip
The solid black lines in Figs~\ref{fig:2}--\ref{fig:3} represent the case of $J_2=0$, i.e., the $1$-NNCRNs without intrinsic correlations at distance $l\geq 2$, as the baseline for comparison.
With the exception of the rightmost panel of Fig~\ref{fig:3}, the peak position of susceptibility suggests that $p_c$ increases with larger $r_2$-values, for fixed $r_1$.

\smallskip As discussed in Sec \ref{sec:rob}, there may be two peaks in susceptibility in the power-law networks with $r_1=0.74$, see the bottom right panel of Fig~\ref{fig:3}.
The first peak appears at $p=0.052$ and is independent of $r_2$, while the second peak appears at larger $p$ and varies with $r_2$.
Moreover, the direction of change (i.e., the qualitative effect of $r_2$ on the location of the peak), is opposite to the behavior mentioned before. Double-peak transitions have been reported in networks with high clustering coefficients \cite{Colomer-de-Simon2014}, and they are known to be characteristic of networks with a core-periphery structure.
The first peak signifies the emergence of a dense core giant component dominated by high-degree nodes, determining the position of $p_c$.
The second peak can be interpreted as the percolation into the periphery of the giant component and is closely related to the size of $R$.

\smallskip
Note moreover that for different values of $r_2$ the relative order of $S(p)$ flips at a certain crossing point, i.e., up to that point the giant connected component with large $r_2$ is larger than the one corresponding to the network with small $r_2$, while the reverse holds above the crossing point.
In particular, for large value of $p$ the giant component size grows more slowly and is smaller with larger $r_2$.
The two regions before and after the crossing point are aggregated into a single number $R$. 
Thus, when discussing the relationship between 2NNCRNs and robustness, it is necessary to look at both measurements, $f_{0.005}$ and $R$.

\smallskip The two measures $f_{0.005}$ and $R$ used to characterize the robustness of $2$-NNCRNs are derived from Figs~\ref{fig:2}--\ref{fig:3} and the result is summarized in Fig~\ref{fig:4}. We observe that $f_{0.005}$ increases with $r_2$ (or remains constant in the case of assortative power-law networks), which means that the effect of perturbing $r_2$ in the positive/negative direction away from $r_2^{\operatorname{ext}}$ is to make the network more robust/fragile.
Conversely, $R$ tends to decrease with increasing $r_2$, which means that $r_2$ has a negative effect on robustness quantified by $R$.
Quantitatively, the size of the shift in the robustness measures resulting from a change in $r_2$ is comparable to the effect of a change in $r_1$, as explained in the caption of Fig~\ref{fig:4}.
The shift induced by $r_2$ is especially large in cases where the range of $r_2$ is wide, such as Poisson networks with $r_1=0.00$ or power-law networks with $r_1=-0.74$. 

\smallskip Unlike most other cases, no change in $f_{0.005}$ due to $r_2$ is observed in power-law networks with a strongly positive $r_1$.
This is because strong positive correlations under high degree fluctuations (with $\gamma=2.5$) induce a core-structure in networks.
For any positive $p$, a component of macroscopic size can be created by connections within the core and to the neighborhood of the core, hence $p_c$ is pushed towards $0$ and $f_c$ towards $1$ as the network size grows.
The invariance of $f_{0.005}$ suggests the change in $r_2$ due to our algorithm cannot dismantle this core.

\smallskip We have also performed the same analysis for random node failure, instead of edge failure. The results are similar and are summarized in Fig~\ref{fig:4_site}.

\subsection{Targeted attack}
It is known that networks that are robust against random failures can still be vulnerable to targeted attacks, where nodes are removed in order of decreasing degree, in particular in random networks with power-law degree distribution \cite{Albert2001,Cohen2001}.
On the other hand, when robustness is evaluated by $f_c$, it is possible to be robust against both random failure and targeted attacks in networks with positive $r_1$~\cite{Tanizawa2012}.
Here, we investigate how the $r_2$-dependence of robustness against targeted attacks differs from the case of random failure.

\begin{figure}[h]
    \centering
     \includegraphics[trim={0cm 0cm 0cm .7cm}, clip, width=1.0\linewidth]{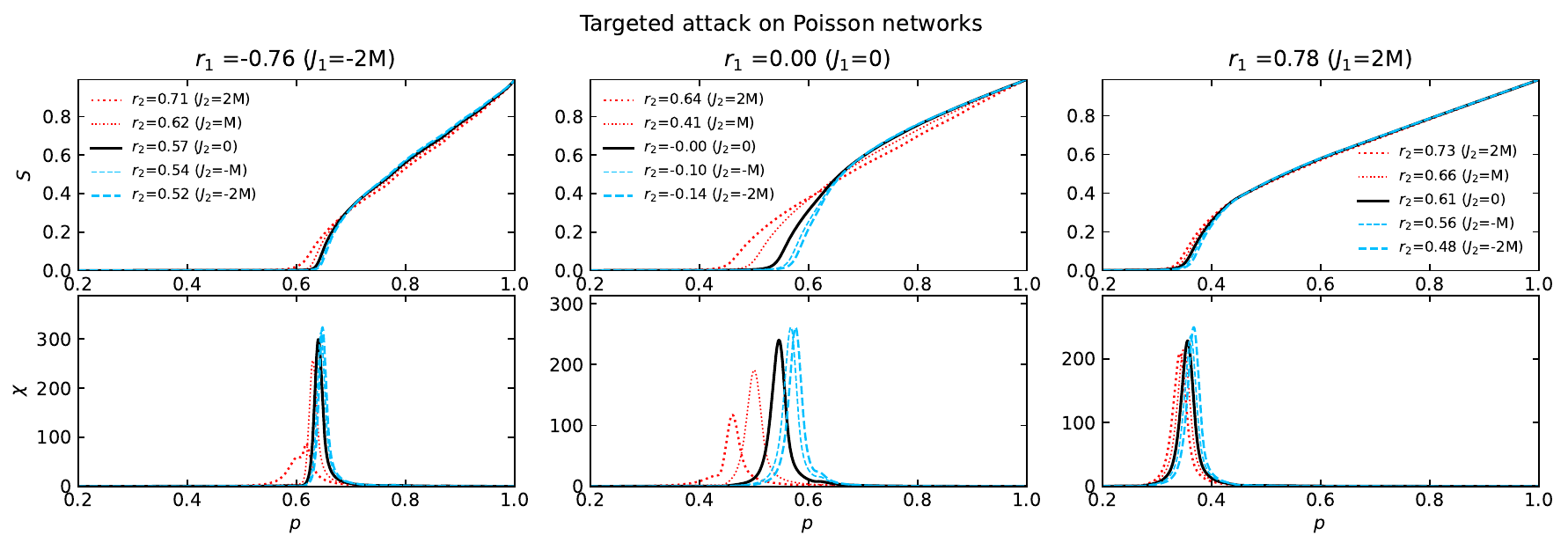}
    \caption{This plot replicates the information from Fig~\ref{fig:2} (Poisson networks) for the case of targeted attack.}
   \label{fig:5}
 \end{figure}

\begin{figure}
     \includegraphics[trim={0cm 0cm 0cm .7cm}, clip, width=1.0\linewidth]{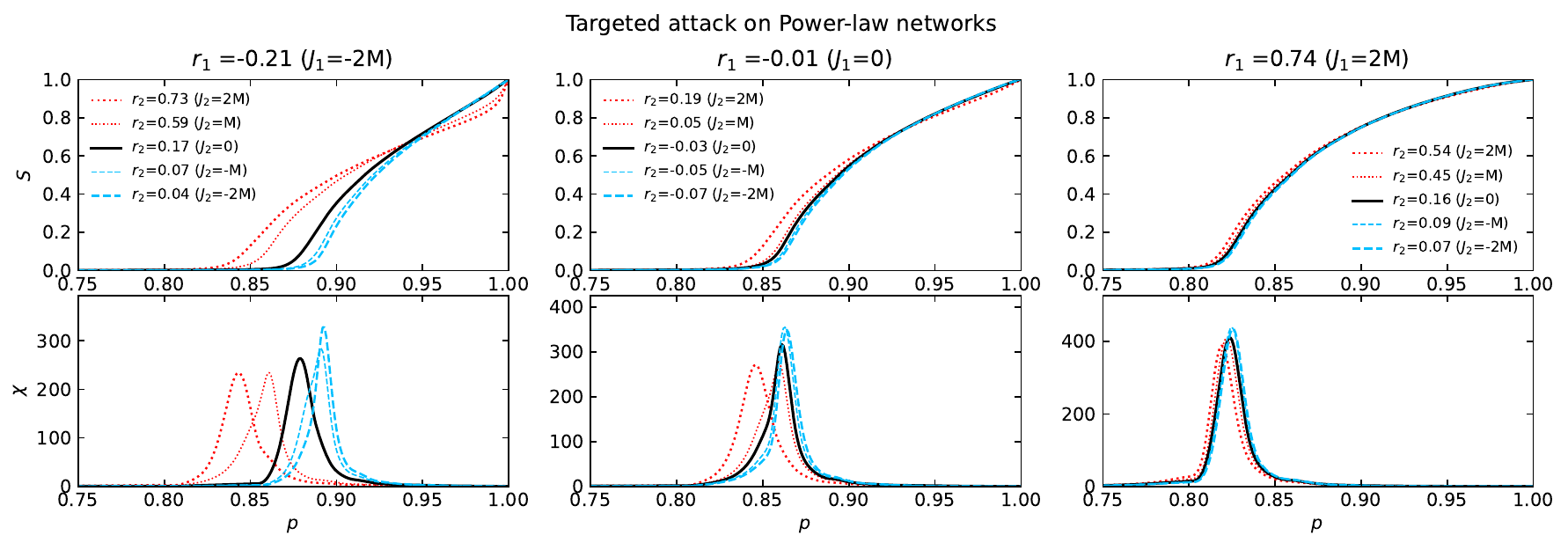}
    \caption{This plot replicates the information from Fig~\ref{fig:3} (power-law networks) for the case of targeted attack.}
    \label{fig:6}
\end{figure}

\begin{figure}[h]
    \centering
    \begin{subfigure}[b]{0.4\linewidth}
        \centering
        \includegraphics[trim={0cm 0cm 0cm 1.3cm}, clip, width=\linewidth]{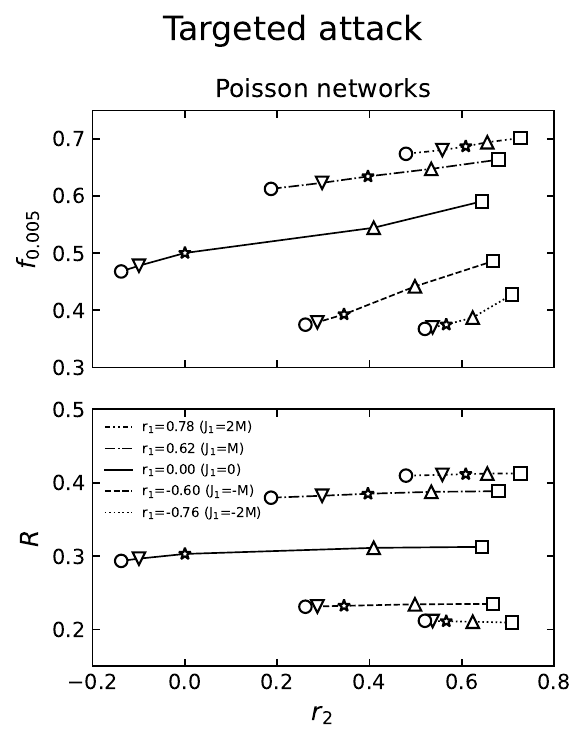}
    \end{subfigure}%
    \begin{subfigure}[b]{0.4\linewidth}
        \centering
        \includegraphics[trim={0cm 0cm 0cm 1.3cm}, clip,  width=\linewidth]{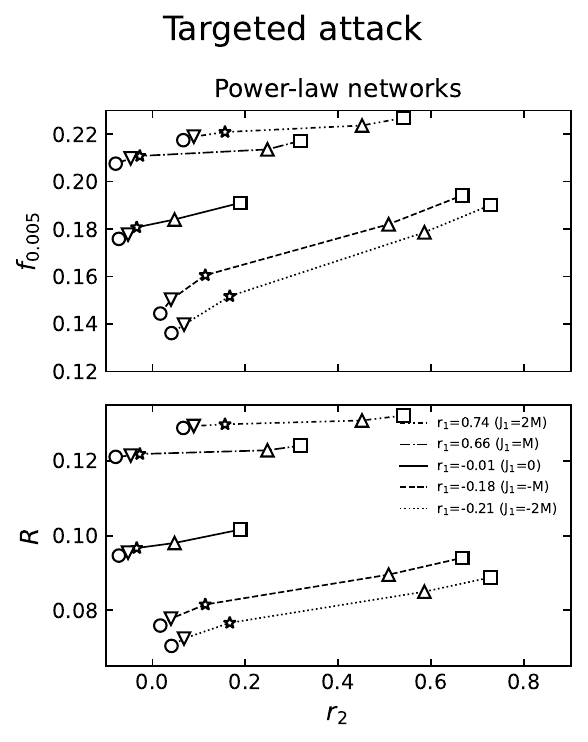}
    \end{subfigure}
    \caption{ Robustness of $2$-NNCRNs against targeted attack, analogously to Figs~\ref{fig:4}--\ref{fig:4_site}.}
    \label{fig:7}
\end{figure}

\smallskip Figs~\ref{fig:5}--\ref{fig:6} depict how the giant component size $S$ and the susceptibility $\chi$ depend on the fraction $p$ of surviving nodes under a targeted attack on the same $2$-NNCRNs used in Figs~\ref{fig:2}--\ref{fig:3}. Fig~\ref{fig:7} contains the corresponding values of $f_{0.005}$ and $R$.

\smallskip We observe that $f_{0.005}$ increases with $r_2$ for a fixed choice of $r_1$ and that this behavior qualitatively matches our observation in the case of random edge failure. Quantitatively, the strength of the shift in $f_{0.005}$ in response to a change in $r_2$ is stronger than in the case of random failure. Moreover, as in the case of random failure, the size of the shift in response to a change in $r_2$ is somewhat weaker compared to a change in $r_1$.

\smallskip On the other hand, as observed from the bottom panels in Fig~\ref{fig:7}, there is a slight increase in $R$ with respect to $r_2$, which is more pronounced in power-law networks. We note that the direction of change in $R$ in response is opposite to what we observed in random failure.
This difference in behavior is due to the fact that, while there is still a crossing point between $S(p)$ similar to the discussion in Sec.~\ref{randfail}, it occurs at a higher proportion $S(p)$ of surviving nodes than in the case of random failure. Thus the values of $R$ are influenced more by the behavior of the curves around $p_c$ and we expect that the change in $R$ more closely matches the change in $f_{0.005}$.

\section{Conclusion and Discussion}\label{sec:end}
In this study, we have conducted numerical investigations into the structural properties of networks exhibiting degree-degree correlations between second-nearest neighbors, i.e., long-range degree correlations (LRDCs) at shortest path distance $l=2$.

To achieve this, we have introduced $l$-th nearest-neighbor correlated random networks ($l$-NNCRNs), which exhibit degree-degree correlations up to the $l$-th nearest neighbor scale while being maximally random at any further scale.
We have generated $2$-NNCRNs using a two-step algorithm based on random edge rewiring and the Metropolis-Hastings algorithm for two representative degree sequences: Poisson and power-law distributions.
Quantifying the strength of the LRDC at distance $l$ using Pearson's correlation coefficient $r_l$, this two-step algorithm generates $2$-NNCRNs with identical $r_1$ but varying $r_2$.
Finally, we have investigated the robustness of the networks by simulating random node/edge failures and degree-based targeted attacks.

\smallskip
We have observed that the robustness quantified by $f_{0.005}$ is an increasing function of $r_2$ when $r_1$ is held constant in the all cases we simulated (Figs~\ref{fig:4}, \ref{fig:4_site} and \ref{fig:7}). Thus, networks with smaller (larger) $r_2$ than $r_2^{\operatorname{ext}}$ have been observed to be more fragile (more robust) against both random failure and targeted attack.
On the other hand, the effect of $r_2$ on network robustness, when measured by $R$, is more nuanced. In the case of bond percolation (Fig~\ref{fig:4}), $R$ decreases as $r_2$ exceeds $r_2^{\operatorname{ext}}$, indicating that the network becomes more fragile. However, in the case of targeted attacks (Fig~\ref{fig:7}), $R$ remains constant or increases slightly with $r_2$. This contrasts with our findings when robustness is quantified by $f_{0.005}$ and can be explained by the crosspoint-effect mentioned in Section~\ref{randfail}.

\smallskip To summarize, our numerical results suggest that network robustness is correlated with $r_2$ when $r_1$ is fixed, as shown in Table~\ref{fig:1}. The trends in the influence of $r_2$ on robustness are similar to those of $r_1$. As seen in Figs~\ref{fig:4}, \ref{fig:4_site}, and \ref{fig:7}, the magnitude of the effect caused by changes in $r_2$ is roughly half of that caused by $r_1$, except in the case of $R$ during targeted attacks. This highlights the significant role of second-neighbor degree-degree correlations in network resilience.
\begin{table}[b]
\begin{tabular}{cccc}
\hline\hline
       & $p_{0.005}$ & $f_{0.005}$ & $R$ \\ \hline
Random node/edge failure  & $\downarrow$               & $\uparrow$               & $\downarrow$\\ \hline
Degree-based target attack  & $\downarrow$               & $\uparrow$               & $\uparrow$\\ \hline\hline
\end{tabular}
\caption{Summary of effects of an increase in either $r_1$ or $r_2$ on $p_{0.005}$, $f_{0.005}$, and R.}
\end{table}

\begin{figure}
    \centering
     \includegraphics[width=0.5\linewidth]{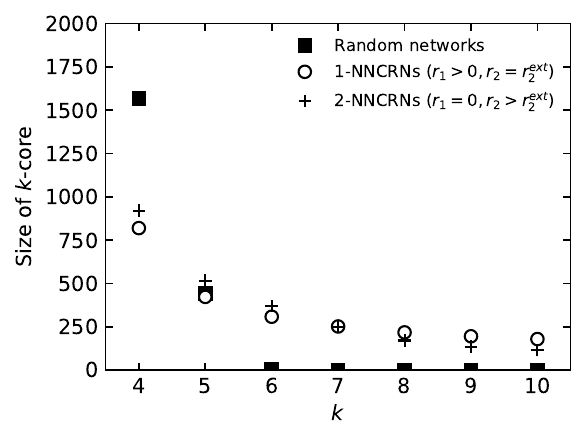}
    \caption{Size of $k$-cores in power-law networks with different degree correlations. The values are averaged over $100$ configurations. Symbols represent different rewiring parameters: solid squares for $(J_1, J_2) = (0, 0)$, open spheres for $(J_1, J_2) = (2M, 0)$, and crosses for $(J_1, J_2) = (0, 2M)$, which means that the networks are random networks, $1$-NNCRNs, and $2$-NNCRNs, respectively.
    }
    \label{fig:8}
\end{figure}

\smallskip In conclusion, if $r_1$ is kept fixed, then $r_2$ has a similar effect on robustness as $r_1$. To give some explanation of this effect, we have plotted in Fig~\ref{fig:8} the size of the $k$-core in power-law networks with strongly positive $r_1$ or $r_2$ and we observe that both constraints lead to quite similar values. To interpret this, recall that the $k$-core is the part of the network where the giant component starts to emerge as $p$ approaches $p_c$ from below and is thus intimately related to the robustness of the network. Quantitatively, the effect of $r_2$ on robustness is smaller than $r_1$ but comparable. 

\begin{figure}
    \centering
     \includegraphics[width=0.75\linewidth]{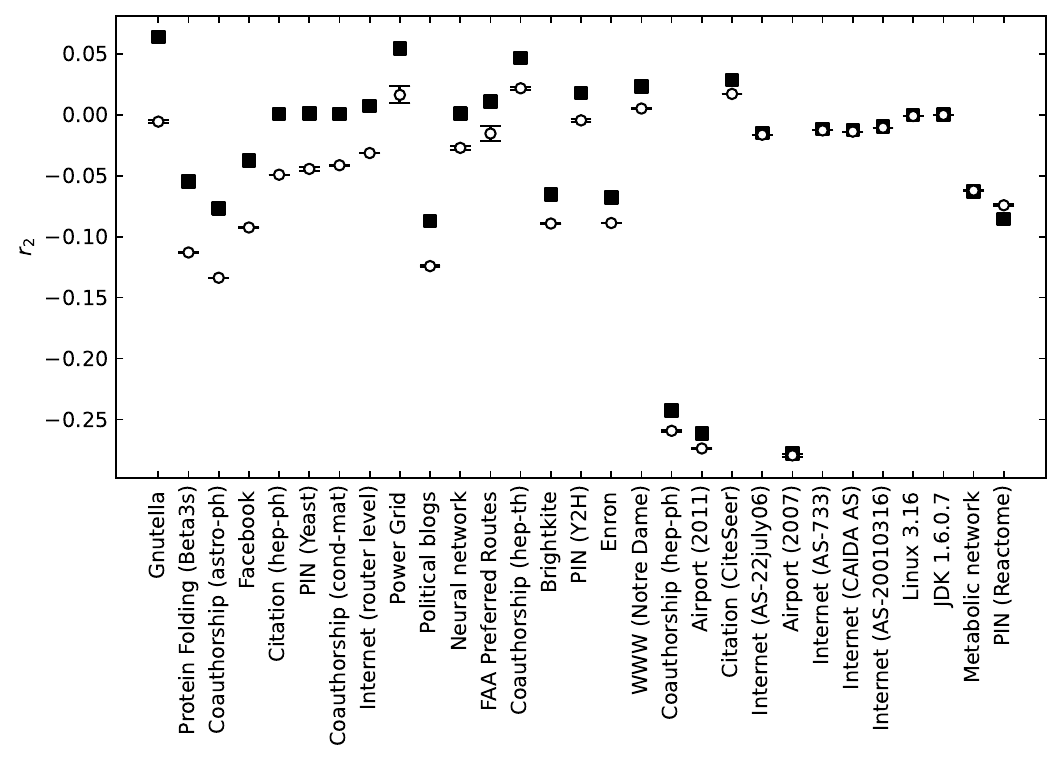} 

    \caption{LRDCs at $l=2$ in real networks in Table~\ref{tab:2}. Solid squares represent the $r_2$-values for real-world networks. Open circles denote $r_2^{\operatorname{ext}}$ calculated as the average $r_2$ values over $100$ realizations of the corresponding $1$-NNCRNs, with error bars indicating the standard deviation.}
    \label{fig:9}
\end{figure}

\smallskip It is natural to expect that LRDCs at $l=2$ also play an important role in the robustness of a real-world network. Indeed, Fig~\ref{fig:9} summarizes the observed values of $r_2$ for real-world networks and the value of $r_2^{\operatorname{ext}}$ for a $1$-NNCRN with the same $P(k,k'|l=1)$. Overall, real-world networks tend to have larger $r_2$ than $r_2^{\operatorname{ext}}$, and thus our results suggest that intrinsic LRDCs at $l = 2$ in real-world networks contribute positively to structural robustness. It remains an important problem for future work to quantify the influence of LRDCs in such networks.
\begin{table}
\begin{tabular}{llll}
\hline\hline
Network & N & M & Ref. \\
\hline
Airport (2007) & 500  & 2,980  & \cite{Latora2017,Colizza2007} \\
Airport (2011) & 1,574  & 17,215  & \cite{Opsahl2011} \\
Brightkite  & 58,228  & 214,078  & \cite{snapnets,Cho2011} \\
Citation (CiteSeer) & 384,054  & 1,736,145  & \cite{peixoto,Bollacker1998} \\
Citation (hep-ph) & 34,546  & 420,877  & \cite{snapnets,Leskovec2005, Gehrke2003} \\
Coauthorship (astro-ph) & 18,771  & 198,050  & \cite{snapnets,Leskovec2007} \\
Coauthorship (cond-mat) & 23,133  & 93,439  & \cite{snapnets,Leskovec2007} \\
Coauthorship (hep-ph) & 12,006  & 118,489  & \cite{snapnets,Leskovec2007} \\
Coauthorship (hep-th) & 9,875  & 25,973  & \cite{snapnets,Leskovec2007} \\
Enron  & 36,692  & 183,831  & \cite{snapnets,Leskovec2009, Klimt2004} \\
FAA Preferred Routes  & 1,226  & 2,408  & \cite{peixoto,FAA2010} \\
Facebook  & 63,731  & 817,035  & \cite{Viswanath2009} \\
Gnutella  & 10,876  & 39,994  & \cite{snapnets,Leskovec2007, Ripeanu2002} \\
Internet (AS-20010316) & 10,515  & 21,455  & \cite{cosinproject,cosinproject} \\
Internet (AS-22july06) & 22,963  & 48,436  & \cite{newman} \\
Internet (AS-733) & 6,474  & 12,572  & \cite{snapnets,Leskovec2005} \\
Internet (CAIDA AS) & 26,475  & 53,381  & \cite{snapnets,Leskovec2005} \\
Internet (router level) & 192,244  & 609,066  & \cite{caida} \\
JDK 1.6.0.7  & 6,434  & 53,658  & \cite{peixoto,Kunegis2013} \\
Linux 3.16  & 30,834  & 213,217  & \cite{peixoto,Kunegis2013} \\
Metabolic network  & 453  & 2,025  & \cite{arenas,Duch2005} \\
Neural network  & 297  & 2,148  & \cite{newman,White1986, Watts1998} \\
PIN (Reactome) & 6,229  & 146,160  & \cite{peixoto,Joshi-Tope2005} \\
PIN (Y2H) & 3,023  & 6,149  & \cite{peixoto,Rual2005} \\
PIN (Yeast) & 2,284  & 6,646  & \cite{pajek,Bu2003} \\
Political blogs  & 1,224  & 16,715  & \cite{newman,Adamic2005} \\
Power Grid  & 4,941  & 6,594  & \cite{newman,Watts1998} \\
Protein Folding (Beta3s) & 132,167  & 228,967  & \cite{cosinproject,Rao2004} \\
WWW (Notre Dame) & 325,729  & 1,090,108  & \cite{snapnets,Albert1999} \\
\hline\hline
\end{tabular}
\caption{Properties of the real networks in Fig~\ref{fig:9}}
\label{tab:2}
\end{table}

\smallskip The insights gained from our study have important implications for the construction of more robust social systems and the development of effective contact rules to mitigate the spread of infectious diseases. Our results show that $r_2$ significantly affects network robustness, with its positive impact explained by the $k$-core structure induced when $r_2 > r_2^{\operatorname{ext}}$. Given that nonzero $r_1$ has emerged as a fundamental property of network structure, influencing not only network robustness but also various dynamic processes and functions, it is likely that $r_2$ plays a similar role. Further investigations are needed to explore the influence of $r_2$ on other dynamic processes.

\begin{acknowledgments}
The authors thank S.~Mizutaka, T.~Hasegawa, and K.~Yakubo for fruitful discussions.
This work was supported by a Grant-in-Aid for Scientific Research (No.~23K13010 and 23K12984) from the Japan Society for the Promotion of Science.
\end{acknowledgments}

\bibliographystyle{apsrev4-2}
\bibliography{ref}
\printindex

\end{document}